\documentclass[a4paper,fleqn]{cas-dc}

\usepackage[square,sort,comma,numbers]{natbib}

\def\tsc#1{\csdef{#1}{\textsc{\lowercase{#1}}\xspace}}
\tsc{WGM}
\tsc{QE}

\usepackage{microtype}
\usepackage{graphicx}
\usepackage{subfigure}
\usepackage{booktabs} 
\usepackage{mathtools}
\usepackage{amsmath,amsfonts}
\usepackage{algorithm}
\usepackage{algorithmic}
\usepackage{hyperref}

\newcommand{\TheName}{HeterPS}

\newcommand{\rev}[1]{{\color{black}{#1}}}
\newcommand{\revv}[1]{{\color{black}{#1}}}

\begin{document}

\let\WriteBookmarks\relax
\def\floatpagepagefraction{1}
\def\textpagefraction{.001}

\shorttitle{Distributed Deep Learning With Reinforcement Learning Based Scheduling in Heterogeneous Environments} 

\shortauthors{Liu et al.} 

\title [mode = title]{\TheName{}: Distributed Deep Learning With Reinforcement Learning Based Scheduling in Heterogeneous Environments}

\author[1]{Ji Liu}[]\fnmark[$\dagger$]\cormark[1]
\author[2]{Zhihua Wu}[]\fnmark[$\dagger$]
\author[1]{Danlei Feng}[]
\author[1]{Minxu Zhang}[]
\author[1]{Xinxuan Wu}[]
\author[1]{Xuefeng Yao}[]
\author[1]{Dianhai Yu}[]
\author[1]{Yanjun Ma}[]
\author[2]{Feng Zhao}[]\cormark[1]
\author[1]{Dejing Dou}[]

\affiliation[1]{organization={Baidu Inc.},country={China}}

\affiliation[2]{organization={University of Science and Technology of China},country={China}}

\cortext[1]{Corresponding author\\\indent \indent $\dagger$Equal contribution}

\tnotemark[1] 


\begin{abstract}
Deep neural networks (DNNs) exploit many layers and a large number of parameters to achieve excellent performance. The training process of DNN models generally handles large-scale input data with many sparse features, which incurs high Input/Output (IO) cost, while some layers are compute-intensive. The training process generally exploits distributed computing resources to reduce training time. 
While heterogeneous computing resources, e.g., CPUs, GPUs of multiple types, are available for the distributed training process, the scheduling of multiple layers to diverse computing resources remains critical for the training process. To efficiently train a DNN model using the heterogeneous computing resources, we propose a distributed framework, i.e., \underline{Heter}ogeneous \underline{P}arameter \underline{S}erver (\TheName{}), composed of a distributed architecture and a Reinforcement Learning (RL)-based scheduling method. The advantages of \TheName{} are three-fold compared with existing frameworks. 
First, \TheName{} enables efficient training process of diverse workloads with heterogeneous computing resources. 
Second, \TheName{} exploits an RL-based method to efficiently schedule the workload of each layer to appropriate computing resources to minimize the cost while satisfying throughput constraints.
Third, \TheName{} manages data storage and data communication among distributed computing resources.
We carry out extensive experiments to show that \TheName{} significantly outperforms state-of-the-art approaches in terms of throughput (14.5 times higher) and monetary cost (312.3\% smaller). 
\end{abstract}


\maketitle




\section{Introduction}
\label{sec:intro}

Deep Neural Networks (DNNs) have achieved major success in various domains, such as computer vision, natural language processing, and advertising systems.
Models with many layers, neurons and huge amounts of parameters are generally trained with large amounts of data, which drastically improves ultimate accuracy \cite{Dean2012}.
For instance, 
BERT \cite{Devlin19} exploit from 110 million to 340 million parameters, while ERNIE exploits 260 billion parameters \cite{wang2021ernie}. 

Large models are generally composed of both data-intensive and compute-intensive layers. A layer is data-intensive when it takes \rev{much more time for the Input/Output (IO) of the data than that of computation \cite{furht2010handbook}}. Otherwise, the layer is compute-intensive.
For instance, CTR models process high-dimensional input data, which contain a large number of sparse features. Sparse features refer to a small proportion of non-zero features, which are processed through an embedding layer to generate a low-dimensional embedding \cite{Zhao2020Distributed,ArchImpl19}. As the embedding layer processes large amounts of data, e.g., around 10 TB or even bigger, the processing incurs high IO cost and is data-intensive. However, the training process of some other layers in deep neural networks, e.g., fully-connected layers, are compute-intensive because of heavy computing workloads. 

As the processing units, e.g., CPUs, diverse types of GPUs, AI processors, become heterogeneous, it is critical to take full advantage of heterogeneous computing resources for the distributed training process of large-scale DNN models. Some computing resources, e.g., CPUs, \rev{performs well with} data-intensive tasks, while some others, e.g., GPUs, \rev{is suitable for} compute-intensive tasks. 
In this case, the scheduling between tasks and diverse computing resources is of much importance for distributed training. While the scheduling problem is a typical NP-hard problem \cite{Du1989NPHard, liu2020job}, some straightforward methods already exist. For instance, as it generally deals with big amounts of data, the first layer can be scheduled to CPUs, while the other layers are scheduled to GPUs \cite{Zhao2019AIBox}. As not all the DNN models have the same structure, this method may not be suitable for diverse types of DNN structures. \rev{For instance, some layers, which are not the first layer, can be scheduled to CPUs as well for smaller cost.} Genetic \cite{barika2019scheduling}, and Greedy \cite{shi2020device} can be directly applied to address the layer scheduling problem\revv{. However,} they may fall into local optimal and correspond to high cost. Although black-box optimization methods, e.g., Bayesian Optimization (BO)-based scheduling method \cite{Dolatnia2016Bayesian}, can also be applied, BO may incur  high costs due to the randomness of the sampling process. 

After scheduling tasks to proper heterogeneous computing resources, parallelism can be achieved to accelerate the training process. Data parallelism \cite{verbraeken2020survey} is widely used to parallelize the training process of large-scale DNN models, while pipeline parallelism \cite{huang2018gpipe, narayanan2019pipedream, fan2021dapple} emerges as a promising approach to address large DNN models. 
With the data parallelism approach, the training data is partitioned as many times as the number of computing resources, and all computing resources subsequently apply the same DNN model to process different chunks of the data sets \cite{verbraeken2020survey}. With the pipeline approach, each computing resource processes the training data with a stage of the DNN model, while the processing of each stage can be parallelized \cite{narayanan2019pipedream}. A stage of the DNN model is composed of several continuous layers, while two different stages may have data dependencies, i.e., the output of a stage is the input of the other stage.
The data parallelism and the pipeline parallelism can be combined to achieve fine-grained parallelism \cite{fan2021dapple, narayanan2019pipedream}. 

While parallelism reduces the training time, multiple computing resources may incur significant monetary costs. The training process generally has a hard throughput limitation to train a model within an acceptable time. Then, it is beneficial to minimize the monetary cost with the throughput limit. As the computing resources can be elastic, i.e., the number of computing resources can scale up or shrink on demand, the elasticity of the computing resources can be exploited to ensure the throughput constraint while minimizing the monetary cost. In this case, it is critical to determine the number of computing resources for the distributed training.

In this paper, we propose \underline{Heter}ogeneous \underline{P}arameter \underline{S}erver (\TheName{}) to enable the distributed training of large-scale models with elastic heterogeneous computing resources. 
\TheName{} is composed of three modules, i.e., a DNN layer scheduling module, a data management module, and a distributed training module. The DNN layer scheduling module generates a scheduling plan and a provisioning plan. The provisioning plan defines the number of computing resources of each type for the distributed training process, and the scheduling plan maps each layer to a proper type of computing resources. \rev{Within a DNN model, multiple layers may have diverse characteristics, e.g., data-intensive or compute-intensive. To reduce the training time, we schedule each layer to a proper type of computing resource and combine several successive layers into a stage, which is scheduled to the same type of computing resources to reduce the time to transfer data. In this way, a scheduling plan is generated. Afterward, we generate a provisioning plan to adjust the number of computing resources of each type to achieve load balance and to reduce the monetary cost while meeting the throughput \revv{constraint}. Please note that the scheduling is carried out at the layer level while the provisioning is performed at the stage level.} The data management module handles the data transfer among different clusters or servers. A cluster is a set of interconnected computing resources \cite{liu2014parallelization}. The distributed training module parallelizes the training process of the model while exploiting the combination of data parallelism and pipeline parallelism. \rev{To the best of our knowledge, we are among the first \cite{jiang2020unified,jia2022whale,park2020hetpipe,narayanan2020heterogeneity} to enable large-scale DNN with elastic heterogeneous computing resources, which can minimize the total cost while ensuring the throughput. }

We summarize our main contributions as follows:

\begin{itemize}
    \item We propose a novel framework denoted by \TheName~to enable the distributed training of large-scale DNN with elastic heterogeneous computing resources. The framework manages the data storage and data communication among distributed computing resources.
    \item We propose a novel Reinforcement Learning (RL)-based layer scheduling method \rev{with an LSTM model} to schedule each layer to a proper type of computing resources while minimizing the total cost and ensuring the throughput. In addition, based on the scheduling method, we \rev{design} a method to determine the proper number of computing resources for distributed training.
    \item We carry out extensive experiments based on DNN models of diverse structures \rev{to validate our proposed approach, which} show the advantages of our approach compared with baseline approaches.
\end{itemize}


The rest of the paper is organized as follows. Section \ref{sec:relatedwork} presents the background and related work. In Section \ref{sec:systemModel}, we introduce the system design of \TheName{}, including the system architecture and the data management methods. 
In Section \ref{sec:problem}, we explain the cost model and formally define the scheduling problem. In Section \ref{sec:solution}, we propose a reinforcement learning-based scheduling method.
Section \ref{sec:experimentation} presents extensive experimental evaluation results to show the advantages of \TheName{}. Finally, Section \ref{sec:con} concludes.

\section{Background \& Related Works}
\label{sec:relatedwork}

In this section, we present the related work for distributed training and scheduling methods. 


\subsection{Distributed Training}



Data parallelism \cite{verbraeken2020survey, Shen2020PyTorch} and pipeline parallelism \cite{huang2018gpipe, narayanan2019pipedream, fan2021dapple}, can be utilized to parallelize the training process. 
The data parallelism approach enables the training process to handle large amounts of data. The input data is partitioned as many times as the number of computing resources, and each part of the data is used to calculate the gradients at each computing resource. The model parameters or gradients at each computing resource are averaged to update the parameters in each computing resource, using either a parameter server architecture \cite{li2014scaling} or a ring-allreduce architecture \cite{ring-allreduce}. 
The parameter server architecture exploits a server to calculate the global parameters or gradients based on a distributed optimization algorithm, while the ring-allreduce architecture exploits a decentralized protocol to calculate the global parameters or gradients.
The pipeline approach partitions the DNN model into multiple stages, while each stage is processed in a computing resource \cite{narayanan2019pipedream}. 
The data parallelism and the pipeline parallelism can be combined to parallelize the execution of each stage \cite{fan2021dapple, narayanan2019pipedream}. 

Although many works \cite{fan2021dapple, narayanan2019pipedream, Huang2019Advances,zheng2022alpa} have been proposed to combine the data parallelism and the pipeline parallelism to accelerate the distributed training, they focus on GPUs and cannot utilize heterogeneous computing resources, e.g., diverse types of GPUs. 
Parameter servers can exploit CPUs to calculate the global parameters or gradients \cite{li2014scaling} while GPUs can also be used in parameter servers to accelerate the processing \cite{cui2016geeps, Zhao2020Distributed}. 
However, these approaches cannot take advantage of the combination of CPUs or GPUs to handle both data-intensive tasks and compute-intensive tasks.
\revv{Alpa \cite{zheng2022alpa} enables automatic inter- and intra-operator parallelism, which does not consider the real dynamic execution information or the overlap between computation and communication. In our work, we consider the real dynamic execution environments and the overlap between the computation and communication in order to achieve excellent performance.}
Some works \cite{kim2019parallax,jiang2019xdl} combine the parameter server architecture (for sparse layers) and ring-allreduce architecture (for dense layers), while they do not focus on heterogeneous computing resources. AIBox \cite{Zhao2019AIBox} and BytePS \cite{jiang2020unified} can exploit CPUs and GPUs while they utilized heuristics to statically schedule tasks to CPUs or GPUs, which cannot take advantage of the elasticity of available computing resources \rev{e.g., in the Cloud, Grid or similar environments}. \revv{Some works, e.g., Whale \cite{jia2022whale}, HetPipe \cite{park2020hetpipe}, and Gavel \cite{narayanan2020heterogeneity}, can enable model training with heterogeneous resources. However, Whale only considers memory constraint without considering the throughput constraint or the costs for the training process. 
HetPipe only considers static profiled information and exploits heuristics for the training process. 
Gavel focuses on scheduling the whole training process of a model to a certain type of resource without partitioning, which corresponds to low throughput and high costs with elastic resources. In our work, we enable the training process with elastic heterogeneous resources while considering the throughput constraint and minimizing the costs.}

\begin{figure}[!t]
\centering
\includegraphics[width=0.49\textwidth]{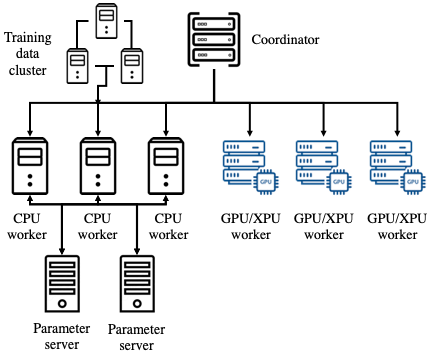}
\caption{The infrastructure architecture of \TheName{}. A coordinator is connected to each computing resource (worker). Each worker can be CPU-based or GPU or XPU-based. XPU represents diverse types of processors optimized for the training process of DNN models. The training data is stored in a cluster, which is also connected to the workers and the coordinator.}
\label{fig:infraArchi} 
\end{figure}

\subsection{Scheduling Method}

Traditional distributed machine learning \cite{li2014scaling, ring-allreduce} is generally carried out with homogeneous computing resources, e.g., CPUs or GPUs. Heterogeneous computing resources are considered to deal with the distributed training process of CTR models in \cite{Zhao2019AIBox}, exploiting a simple heuristic scheduling method that directly schedules the first layer to CPUs and the other layers to GPUs.
Greedy methods \cite{shi2020device} can be utilized to minimize the energy cost for federated edge learning \cite{che2022federated,jin2022accelerated,li2022fedhisyn,liu2022multi,zhang2022fedduap} with CPU-GPU heterogeneous computing environments, which may fall into the local optimal, corresponding to a high cost. Other heuristics, e.g., Genetic \cite{barika2019scheduling}, can be directly applied to perform the scheduling process, which corresponds to low performance, i.e., relatively high cost. Bayesian optimization can be used to generate good scheduling plans, while it may add much randomness (see details in Section \ref{subsec:layerScheduling}) to the scheduling process \cite{shahriari2015taking,muller2021local,peng2022communication,chawla2010power} and the corresponding cost may be high in some corner cases. While it can be exploited to address the scheduling problem, dynamic programming cannot exploit the real execution time to further optimize the scheduling process \cite{hu2021pipeline}. 

Some existing RL-based scheduling methods focus on the training time and static computing resources without considering the monetary cost or elastic computing resources \cite{mirhoseini2018a,Mirhoseini2017Device,Mao2016Resource,Zhu2020Learning}. Some of them schedule distributed training to each device \cite{mirhoseini2018a,Mirhoseini2017Device,Mao2016Resource}, which corresponds to a huge search space. Existing methods \cite{Mao2016Resource,dutreilh2011using} schedule devices for each \rev{layer (job)} or exploit fully connected layers \cite{Mao2016Resource}, which do not consider the interaction among multiple \rev{layers (jobs)}. In addition, current works only consider homogeneous environment \cite{dutreilh2011using,he2021pipetransformer} or whether the training should continue within each computing node \cite{Zhu2020Learning}, which is different from layer scheduling with elastic computing resources. 

\rev{In our work, we distinguish data-intensive layers and compute-intensive layers based on profiling results measured from the execution of forward and we backward processing of each layer, and schedule each layer to proper computing resources in order to reduce monetary cost while meeting the throughput constraint within heterogeneous environments. In addition, we design an LSTM model with proper features of each layer for the scheduling process while considering the interaction among divers layers. Furthermore, we dynamically generate the provisioning plan for each stage with the scheduled type of computing resource while balancing the throughput of each stage and exploiting a Newton method to minimize the monetary cost while meeting the throughput constraint.}

\begin{figure}[!t]
\centering
\includegraphics[width=0.35\textwidth]{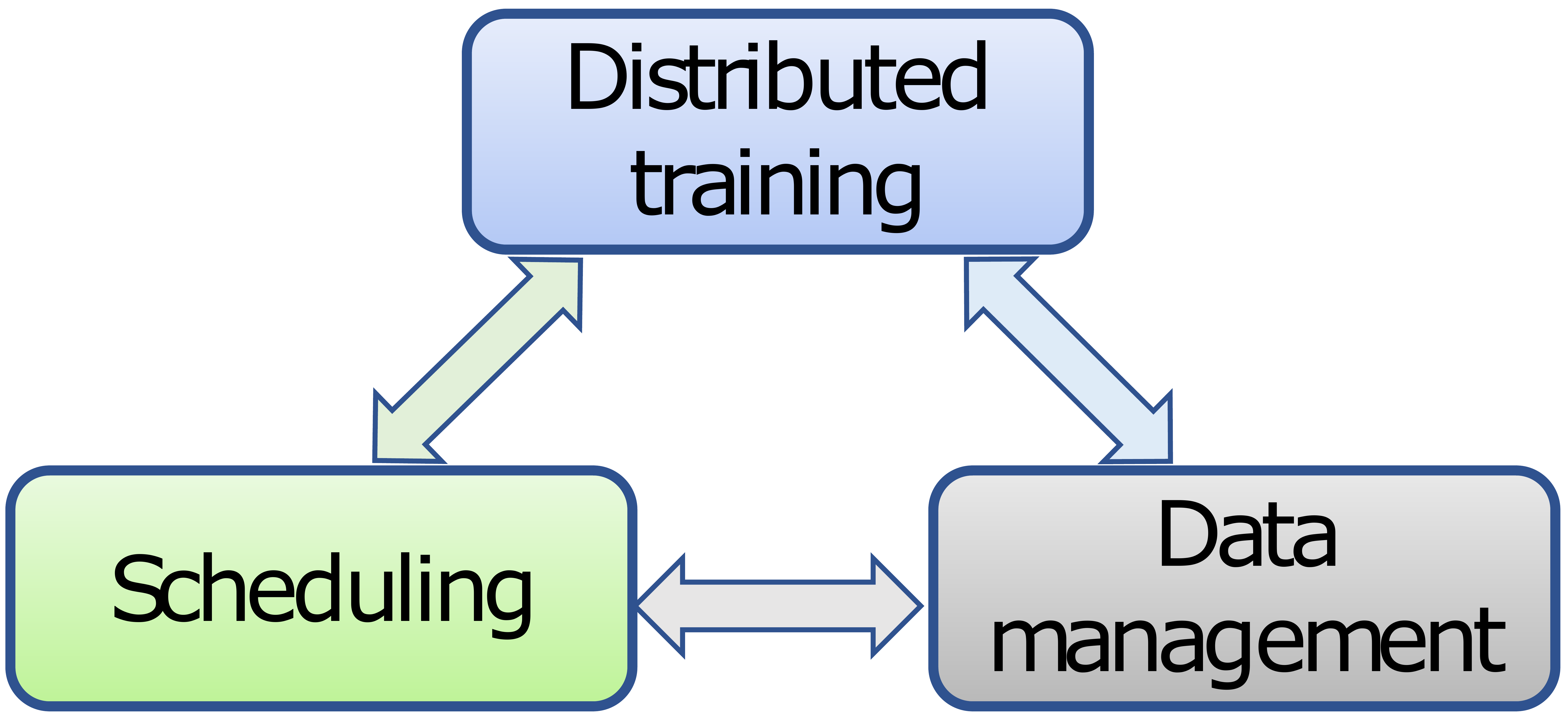}
\caption{The functional architecture of \TheName{}.}
\label{fig:funcArchi} 
\end{figure}

\section{System Design}
\label{sec:systemModel}

In this section, we present the system design of \TheName{}. We first present the system architecture that enables distributed deep learning with heterogeneous computing resources. Then, we present the functionality of each module in \TheName{}, including the distributed training and data management methods.


As shown in Figure \ref{fig:infraArchi}, the infrastructure architecture of \TheName{} consists of four fractions, i.e., training data cluster, coordinator, CPU workers, and GPU/XPU workers. The training data is stored in the training data cluster, e.g., an HDFS cluster \cite{borthakur2008hdfs}. There are three types of computing resources, i.e., CPU workers, parameter servers, and GPU/XPU workers. XPU represents diverse types of processors optimized for the training process of DNN models, e.g., Kunlun AI processors \cite{Ouyang2020}. A coordinator is connected to each worker and the training data cluster. 

As shown in Figure \ref{fig:funcArchi}, the functional architecture of \TheName{} consists of three modules: distributed training, scheduling, and data management. The distributed training module performs the forward propagation, backward propagation, and model update. The scheduling module schedules each layer to a proper type of computing resources. The data management module manages the data storage and data communication. 
This module exploits the host memory, GPU memory, and SSD, to store the data during the training process. 
The distributed training module combines the parameter server architecture and the ring-allreduce architecture to fully exploit heterogeneous computing resources. 
The distributed training module utilizes the scheduling module to place each layer into proper computing resources and the data management module to interact with storage resources or to manage data communication among multiple computing resources.
All three modules are \revv{deployed} in the coordinator, while the training module and the data management module are implemented in each worker. \rev{The coordinator is a centralized module, which generates the scheduling plans and provisioning plans, controls the data management and distributed training process. The training module is implemented in the CPU workers or GPU workers as well to perform the distributed training process. In addition, the data management module enables data caching in the training data center and receives data in the CPU or GPU workers. Please note that this is different from the architecture in \cite{liu2020job} and \cite{jiang2020unified}, which only considers homogeneous resources.}

\begin{table*}[htbp]
\caption{Summary of Main Notations}
\begin{center}
\begin{tabular}{cc}
\toprule
Notation & Definition \\
\hline
\label{tab:summary}
$K$;$\{S_1, S_2, ..., S_K\}$ & The number of stages; a set of stages \\
$OCT$; $ODT$ & The original computation time; the original data communication time \\
$B$; $B_o$ & Batch size; the batch size of a small batch to measure $OCT$ and $ODT$ \\
$CT$; $DT$ & The computation time; the data communication time \\
$k_i$ & The number of computing resources at Iteration $i$ \\
$\alpha_i$; $\beta_i$ & The parts of the workload that can be parallelized for computation and data communication \\
$ET_i$ & The execution time of Iteration $i$ \\
$L$ & The number of epochs \\
$Cost$ & The monetary cost of the training process \\
$p_t$ & The cost to use a computing resource of Type $t$\\
$l$; $a_l$; $L$ & The index of a layer; the scheduling action of Layer $l$; the number of all the layers\\
$T$ & The number of computing resource types \\
$G$ & The total number of the gradients of weights to calculate the expected value\\
\bottomrule
\end{tabular}
\end{center}
\end{table*}

The distributed training module exploits an existing DNN training framework, i.e., PaddlePaddle \cite{Ma2019}, for the training process.
Please note that PaddlePaddle is an industrial platform to train DNN models, while \TheName{} is an architecture within PaddlePaddle for the training process of DNN models in heterogeneous environments.
During the training process, the CPU workers exploit the parameter server architecture, while the same types of GPU/XPU workers take advantage of ring-allreduce architecture, which corresponds to smaller data transfer and balanced workload \cite{kim2019parallax}. In addition, we parallelize the computation and the data communication during the training process. For instance, while a computing resource is performing the forward propagation or backward propagation of a batch of training data, the computed gradients or parameters of another batch of training data can be transferred to other computing resources at the same time.

The scheduling module dynamically schedules each layer to proper computing resources based on profiled information and a cost model. To satisfy a predefined throughput constraint, the scheduling module dynamically generates a provisioning plan and a scheduling plan (see details in Section \ref{sec:solution}), while reducing the total monetary cost based on a cost model (see details in Section \ref{sec:problem}). 

The data management module manages both data storage and data communication. As it takes much time to transfer huge amounts of the training data from the training data cluster to CPU workers, \TheName{} prefetches some input training data and caches them in the memory of CPU workers. When the storage space of CPU workers is limited, the data can also be cached in SSDs or hard disks of CPU workers. During the training process, there is a monitor that counts the access (read or write) frequency of each parameter. If the access frequency is high, the monitor marks the parameters as hot parameters, and the data management module dynamically adjusts it \rev{to} the high-speed storage devices \cite{liu2018efficient}, e.g., the host memory of CPU workers or the GPU memory of GPU workers. Otherwise, the monitor marks the parameters as cold parameters, and the data management module puts it to SSDs or normal hard disks. To accelerate the data communication among multiple workers, the data management module dynamically aggregates the data to send to reduce the overhead of the data communication. Please note that we also exploit data compression during the data communication in the data management module.

\section{Formulation}
\label{sec:problem}

In this section, we present a cost model to estimate the throughput and the monetary cost to carry out the distributed training. Then, we formally formulate the scheduling problem to address.

\subsection{Cost Model}
\label{subsec:costModel}

We propose a cost model to estimate the throughput and the monetary cost to train a DNN model with a specific scheduling plan and a provisioning plan.
The cost model is implemented in the scheduling module and under a specific execution environment, e.g., with distributed heterogeneous computing resources.

Let us assume that a DNN model is partitioned to $K$ stages ($\{S_1, S_2, ..., S_K\}$), each of which is composed of sequential layers scheduled to the same type of computing resources based on a scheduling plan. The computing resources are set up based on a provisioning plan. We assume that the computing resources are homogeneous for a stage, while they can be heterogeneous for different stages. 
In addition, we assume that we have the profiling information of each Stage $S_i$ with the computing resource of a single unit (one CPU core or one GPU) and a small batch size of $B_o$, e.g., the Original Computation Time ($OCT_i$) and the Original Time for Data Communication ($ODT_i$) from the current stage to the next stage. \rev{Please note that $OCT_i$ captures the impact of allreduce or parameter server.} $i$ represents the index of the stage. The batch size refers to the number of training examples utilized in one batch. The batch size also equals the number of samples \rev{processed for one update of the model}. The whole training process is composed of multiple epochs, each of which contains a certain fixed number of batches. For instance, when we have $M$ training examples in total and the batch size is $B$, the number of batches for each epoch is $\frac{M}{B}$. The profiling information can be easily obtained by launching the training process on a single server with limited resources. 
Then, for each stage, we can estimate the computation time ($CT_i$) and data communication time ($DT$) of the stage $S_i$ for a single iteration with $k_i$ computing resources using the following formulas based on the Amdahl's law \cite{Sun2010}: 
\begin{equation}
\label{eq:comptime}
CT_i = \frac{OCT_i}{B_o} * (1 - \alpha_i + \frac{\alpha_i}{k_i})
\end{equation}
\begin{equation}
DT_i = \frac{ODT_i}{B_o} * (1 - \beta_i + \frac{\beta_i}{k_i}),
\end{equation}
where $\alpha_i$ and $\beta_i$ represent the parts of the workload that can be parallelized for computation and data communication \rev{while the sequential part ($1 - \alpha_i$ and $1 - \beta_i$), e.g., the synchronization, allreduce, or parameter server overhead, cannot be parallelized}. These two parameters can be obtained using different units of computing resources and corresponding execution time as presented in \cite{Liu2016Multi}. 

Then, we can estimate the execution time ($ET_i$) of the stage ($S_i$) with the assumption that the \rev{computation} and the data communication can be overlapped \rev{(we ignore the small stages where the computation and data communication cannot be overlapped)} as shown in the following formula:
\begin{equation}
\label{eq:et}
ET_i = \max\{CT_i, DT_i\}.
\end{equation}
Then, the throughput of the stage is:
\begin{equation}
\label{eq:throughput}
\text{Throughput}_i = \frac{B}{ET_i},
\end{equation}
where $B$ is the batch size, i.e., the total number of examples used for Iteration $i$. 
As the training process is realized using a pipeline method, the throughput is limited by the smallest throughput. The overall throughput of the training process can be estimated as:
\begin{equation}
\text{Throughput} = \min_{i \in \{1, 2, ..., S\}}{\text{Throughput}_i}
\end{equation}
Finally, the total execution time ($ET$) of the training process when the number of the epoch is $L$ is:
\begin{equation}
ET = L * \frac{M}{\text{Throughput}}.
\end{equation}
In addition, the monetary cost can be estimated as:
\begin{equation}
\label{eq:monetarycost}
\text{Cost} = ET * \sum_{i = t}^{T} p_t * k_t,
\end{equation}
where $p_i$ represents the cost to utilize the computing resource, which equals the price of the computing resource, e.g., the price of using a V100 GPU per minute; $k_t$ represents the number of computing resources of Type $t$; and $T$ is the number of computing resource types. $p_t$ and $k_t$ are specified in the schedule plan $SP$. Please see the summary of notations in Table \ref{tab:summary}.

\subsection{Problem Formulation}

\begin{figure*}[!ht]
\centering
\includegraphics[width=0.65\textwidth]{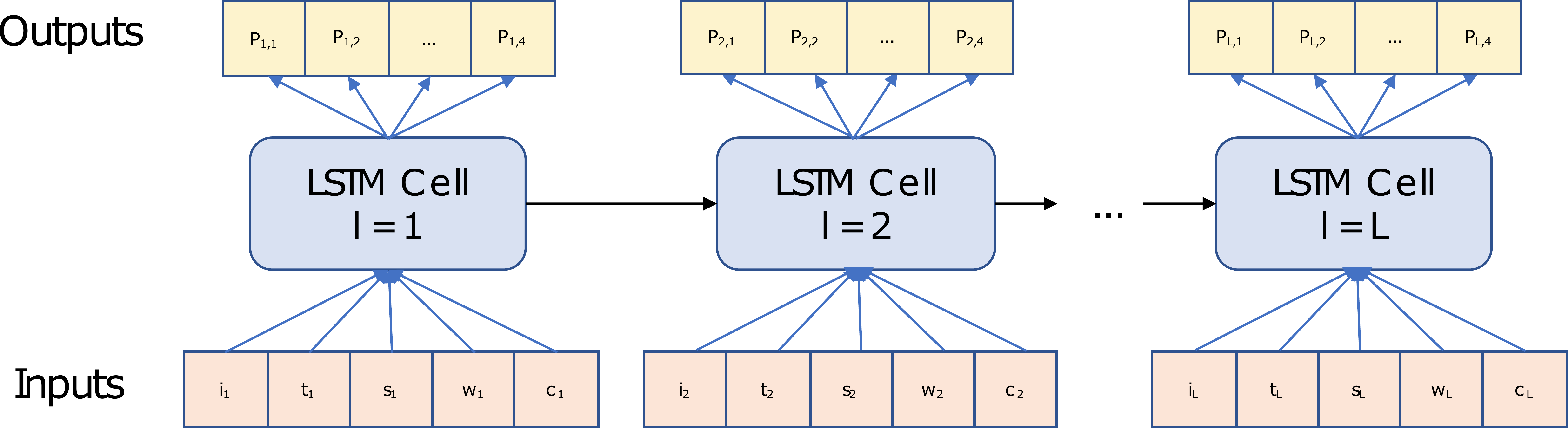}
\caption{The architecture of LSTM. ``i'' represents the index of the layer. ``t'' represents the type of the layer. ``s'' represents the size of the input data of the layer. ``w'' represents the size of the weights of the layer. ``c'' represents the size of data communication time. $P_{l,t}$ represents the probability to schedule Layer $l$ to Type $t$. }
\label{fig:LSTM} 
\end{figure*}

In this section, we formulate the problem of scheduling each layer to a type of computing resource, while minimizing the monetary cost and satisfying the throughput constraint. 
The scheduling process generates a scheduling plan, which is composed of a matrix of decision variables, and each decision variable is defined as follows:
\begin{equation}
\text{Schedule}(l,t) = \begin{cases}
1 & \text{Layer $l$ is scheduled to Type $t$} \\
0 & \text{Otherwise.}
\end{cases}
\end{equation}
A layer can only be scheduled to one type of computing resource. 
The consecutive layers that are scheduled to the same type of computing resources construct a stage. When there are $T$ types of computing resources, each layer can be scheduled to $T$ types. 
The problem we address in this paper is defined as:
\begin{align}
\label{eq:objective}
\min_{SP \in S} & ~ \text{Cost}(SP) \\
s.t. & \begin{cases}
Throughput(SP) > Throughput_{\text{limit}} \label{eq:limit}\\
\text{N}_t(SP) \leq \text{N}_{t, \text{limit}},
\end{cases}
\end{align}
where $Throughput_{limit}$ represents the minimum throughput constraint, and $N_{t, limit}$ represents the maximum number of computing resources for Type $y$. The size of the search space of the scheduling plan is $T^L$ with $L$ representing the number of layers in a DNN model. It is obvious that the search space increases exponentially with $L$. The scheduling problem is a typical NP-hard problem \cite{Du1989NPHard, liu2020job,yu2021toward}.

\section{Reinforcement Learning-Based Scheduling}
\label{sec:solution}

In this section, we first present a method to generate a provisioning plan for each stage to achieve loading \revv{balancing} with a given scheduling plan. Then, we present a Reinforcement Learning(RL)-based scheduling method to generate a scheduling plan.

\subsection{Provision for Load Balancing}
To fully exploit the computing resources for each stage, we try to achieve load balancing for the execution among multiple stages. \rev{Load balancing is achieved in terms of throughput of each stage, which consists of both computing and the data communication of each stage. In this way, we can avoid possible straggler at the stage level of the system.} We choose a proper number of computing resources for each stage, so that the execution time of a single iteration of each stage is similar. When the batch size is small, the execution is communication-intensive, and the execution time of a stage equals the communication time of the data. When the batch size is big enough, the execution is computation-intensive, i.e., the execution time of a stage equals the computation time. \rev{Please note that the distributed training for a stage is based on data parallelism. Then, the distributed training for different stages are achieved by pipeline parallelism or model parallelism. }

Let us illustrate the method to achieve loading balancing given a scheduling plan $SP$ for the case when the batch size is big. In this case, we assume that we can find $\{k_1, k_2, ..., k_n\}$ so that the throughput of each stage is the same. Then, we can have:
\begin{equation}
\text{Throughput}_i = \text{Throughput}_1, \forall i \in \{2, 3, ..., L\}
\end{equation}
Based on Formulas \ref{eq:comptime}, \ref{eq:et}, \ref{eq:throughput}, we can find the relationship between $k_i$ and $k_1$ as shown in the following formula:
\begin{equation}
\label{eq:relation}
k_i = \frac{\alpha_i}{\frac{OCT_1}{OCT_i} * (1 - \alpha_1 + \frac{\alpha_1}{k_1}) - (1 - \alpha_i)}.
\end{equation}
Based on the constraints defined in Formula \ref{eq:limit}, we can have:
\begin{equation}
\label{eq:minimum}
\begin{split}
k_1 > \min \{\frac{\alpha_1 * OCT_1}{Throughput_{limit} * B_o - (1 - \alpha_1) * OCT_1}, \\
\frac{\beta_1 * ODT_1}{Throughput_{limit} * B_o - (1 - \beta_1) * ODT_1}\},
\end{split}
\end{equation}
for the throughput constraint.

In practice, we find the monetary cost increases when $k_1$ increases from the value calculated based on Formula \ref{eq:minimum}. $k_1$ has a limit because the number of the computing resources of Type $1$ is limited as defined in Formula \ref{eq:limit}. We use the Newton method \cite{Hansen1978Newtons} to find a maximum value of $k_1$, which minimizes the monetary cost of the training defined in Formula \ref{eq:objective}, while satisfying Formula \ref{eq:limit}. Then, the value of $k_i$ for each $i \in \{2, ..., n\}$ can be calculated using Formula \ref{eq:relation}. 
In addition, as we exploit the parameter server architecture for the distributed learning, we add an appropriate number of CPU cores to perform the functionality of parameter servers, based on historical profiling results. 
Finally, we can have a provisioning plan for the training with a specific scheduling plan. 

\subsection{Reinforcement Learning-Based Scheduling}

In this section, we present our reinforcement learning (RL)-\cite{Williams1992Reinforce} based scheduling method. As it can well capture the influence of the scheduling decisions of different layers, we exploit a Long Short-Term Memory (LSTM) model to generate a scheduling plan \rev{because of its superior performance} \cite{Greff2017LSTM}. Then, we present the RL-based method to train the LSTM model. 

\begin{algorithm}[!t]
    \caption{Reinforcement Learning-Based Training}
    \label{alg:reinforcement-training}
\begin{flushleft}
    {\bf{Input:}}\\
    \hspace*{0.3in}$N:$ The number of scheduling plans used to train the \hspace*{0.55in}network for each round;\\
    \hspace*{0.3in}$I:$ The maximum round of the training process;\\
    \hspace*{0.3in}$\gamma:$ The rate to update the baseline value;\\
    {\bf{Output:}}\\
    \hspace*{0.3in} $\theta:$ Parameters of the LSTM model;
\end{flushleft}
    \begin{algorithmic}[1]
        \STATE $\theta$ $\leftarrow$ randomly initialize the LSTM model, $b\leftarrow0$ \label{line:initialization}
        \FOR{$i \in \{1,2,...,I\}$}
            \STATE $\mathcal{V} \leftarrow$ generate a set of $G$ scheduling plans\label{line:generatePlan}
            \FOR{$\mathcal{SP} \in \mathcal{V}$}
                \STATE $R_n \leftarrow$ Cost($\mathcal{SP}$) \label{line:computeR}
            \ENDFOR
            \STATE Update $\theta$ according to Formula \ref{eq:rlupdate} \label{line:updateTheta}
            \STATE $b$ $\leftarrow$ (1 - $\gamma)$ * $b$ + $\frac{\gamma}{N}$ * $\sum_{n=1}^N R_n$ \label{line:updateB}
        \ENDFOR
   \end{algorithmic}
\end{algorithm}

As shown in Figure \ref{fig:LSTM}, we build an LSTM model \cite{kawakami2008supervised} with a length similar to the number of layers.
Each LSTM cell takes the input features of each layer and generates the scheduling decision (action). In the LSTM model, we take the index of the layer as the time so that the model can well capture the influence of decision actions of diverse layers. The input of each cell consists of the following five features of each layer: 1. the index of the layer (one-hot encoding); 2. the type of the layer, e.g., full connection, embedding, etc; (one-hot encoding); 3. the size of the input data of the layer (float points); 4. the size of the weights of the layer (float points); 5. data communication time (float points).
The output of each LSTM cell is a vector of $T$ dimensions, i.e., the number of types of computing resources. Then we apply the softmax function on the output vector to get the probability that we schedule the layer to each type of computing resources. Finally, we can generate a scheduling decision with the type of computing resources that correspond to the highest probability for each layer using the LSTM model. 

As the action space in the scheduling plan is not continuous, i.e., a layer can be scheduled to Type 0 (CPU) or Type 1 (GPU of a certain type), the reward signal $R_i$ is non-differentiable and thus the real gradient is hard to obtain. We train the LSTM model based on the rules defined in \cite{williams1992simple}:

\begin{equation}
\label{eqn:approx_reinforce}
\nabla_\theta R(\theta) \approx \frac{1}{G} \sum_{g=1}^G (\sum_{t=1}^T \nabla_\theta \log P(a_t|a_{(t-1):1}; \theta)) R_i,
\end{equation}
where $\nabla_\theta R(\theta)$ represents the gradients of the weights of the LSTM model, $\theta$ is the set of parameters in the LSTM model, $R_i$ is the reward at Iteration $i$, $a_l$ is the scheduling action of Layer $l$, and $P(a_l|a_{(l-1):1}; \theta)$ is the value obtained from softmax function of each layer. We take the monetary cost of $i^{th}$ iteration for the training as the reward, which can be calculated based on Formula \ref{eq:monetarycost}.
To approximate the above gradient, we utilize $G$ gradients of the weights to calculate the expected value. $G$ equals the product of multiplying the number of iterations by the batch size.

Although the original approximation in Formula \ref{eqn:approx_reinforce} is unbiased, it has a very high variance. Inspired by \cite{zoph2016neural}, we alleviate this issue by subtracting the reward with a baseline value, denoted by $b$. We take the moving average of all mean rewards from each batch before the current batch as the baseline value based on following formula:
\begin{equation}
\label{eqn:approx_reinforce_baseline}
\nabla_\theta R(\theta) \approx \frac{1}{G} \sum_{g=1}^G (\sum_{t=1}^T \nabla_\theta \log P(a_t|a_{(t-1):1}; \theta)) (R_i - b)
\end{equation}
Then, we update the parameters of the LSTM model using the following formula:
\begin{equation}
    \label{eq:rlupdate}
    \theta^{'} = \theta + \eta * \nabla_\theta R(\theta),
\end{equation}
where $\eta$ represents the learning rate. 

\begin{figure}[!t]
\centering
\includegraphics[width=0.43\textwidth]{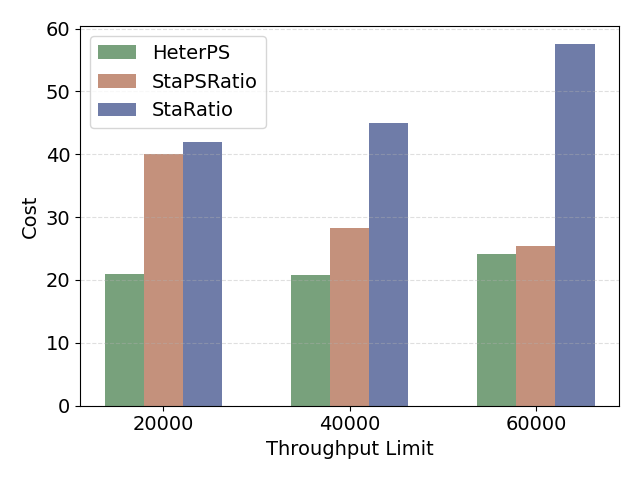}
\caption{The comparison of different methods in terms of monetary cost. The unit of the monetary cost is USD.}
\label{fig:throughputProvisioning} 
\end{figure}

The training process of the LSTM model based on reinforcement learning is shown in Algorithm \ref{alg:reinforcement-training}, which can be utilized in the pre-training process and during the distributed training process. First, the model is initialized (Line 1). With $I$ iterations, the LSTM model is updated (Line 7) using the calculated reward (Line 5) of generated scheduling plans (Line 3). Within the pre-training process, the scheduling plans are randomly generated in Line 3, and the monetary cost is calculated based on Formula \ref{eq:monetarycost}. Within the distributed training process, the scheduling plans are generated based on the updated LSTM model and the monetary cost is calculated based on Formula \ref{eq:monetarycost} with the real throughput. Afterward, the baseline value is updated with the calculated rewards (Line 8). \revv{In Algorithm \ref{alg:reinforcement-training}, the heterogeneity of resources are considered in the scheduling plan, where each layer can be scheduled to a type of resource, and the corresponding cost for the scheduling.} \rev{Please note that Algorithm \ref{alg:reinforcement-training} may also generate a homogeneous scheduling plan, which schedules the same resource type for all layers with the minimum costs while satisfying the throughput constraints, although it generally exploit diverse resources types for the layers with various characteristics.}

\section{Experimental Evaluation}
\label{sec:experimentation}

In this section, we present experimental results to show the advantages of \TheName{}, including the provisioning methods, the RL-based scheduling method, and the throughput of \TheName{} framework. First, we present the comparison to show that our proposed scheduling method outperforms existing provisioning methods. Then, we present the experimental results based on simulation and real execution corresponding to diverse scheduling methods. Please note that aforementioned comparison is carried out within \TheName{} using diverse approaches. Finally, we present the comparison of \TheName{} and Tensorflow \cite{abadi2016tensorflow} in terms of throughput. We implemented \TheName{} using C++ and Python.

Within the experiments, each CPU server has 2 Intel Gold 6271C CPUs, and each CPU contains 24 physical cores. Two NVMe SSDs are attached to each CPU server to support data loading. Each GPU server has the same CPU configurations as that of the CPU servers, while it has 8 Nvidia Tesla V100 GPUs. Both GPU servers and CPU servers are equipped with 512GB Memory. \rev{There are 10 CPU servers and 4 GPU servers in the cluster by default and more servers can be added when required.} All servers are connected with an InfiniBand NIC of 100Gbps bandwidth. The price of each CPU core is 0.04 United States Dollars (USD)/hour, and the price of each GPU card is 2.42 USD/hour. 

\begin{figure}[!t]
\centering
\includegraphics[width=0.48\textwidth]{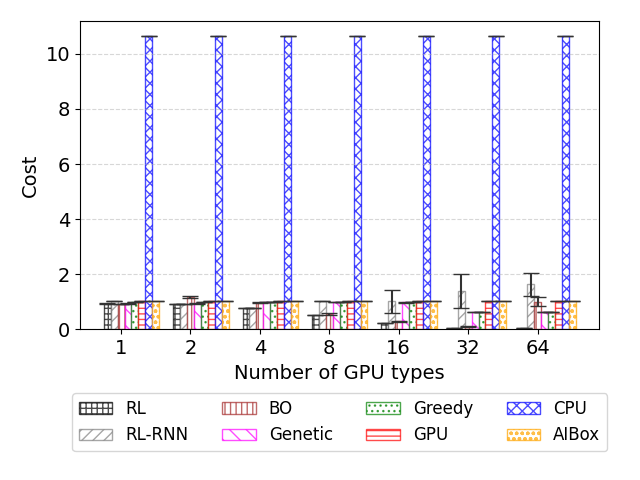}
\vspace{-6mm}
\caption{The cost corresponding to each scheduling method. The cost is normalized by multiplying a constant value for the sake of easy comparison.}
\label{fig:costMatchnetFull} 
\end{figure}

\begin{figure}[!t]
\centering
\includegraphics[width=0.48\textwidth]{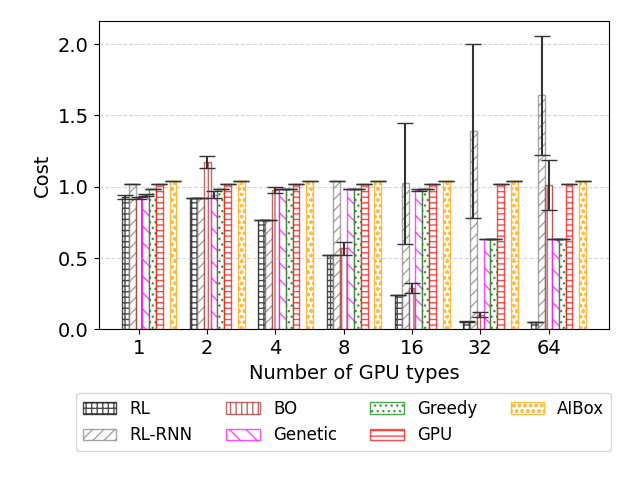}
\vspace{-6mm}
\caption{The cost corresponding to each scheduling method without CPU. The cost is normalized by multiplying a constant value for the sake of easy comparison.}
\label{fig:costMatchnet} 
\end{figure}

\subsection{Computing Resource Provisioning}
\label{subsec:provisioning}

In this section, we present the experimental results by comparing the provisioning plan generated based on our method and static methods, i.e., StaRatio and StaPSRatio. StaRatio represents the ratio between the number of GPU cards and the number of CPU cores is 1:6 \cite{Zhao2019AIBox}, which is the default ratio within a single server. StaPSRatio refers to the ratio with the consideration of parameter servers \cite{jiang2020unified}, i.e., number of GPU cards: number of CPU cores for training: number of CPU cores for parameter servers = 1:6:6, which takes extra CPU cores for parameter servers based on heuristic experience. We obtain the number based on the provisioning method detailed in Section \ref{sec:solution}. In addition, we use the RL scheduling method to perform the scheduling process. 
In this experimentation, we take a CTRDNN model (see details in Appendix)
as a use case.

\begin{figure}[!t]
\centering
\includegraphics[width=0.48\textwidth]{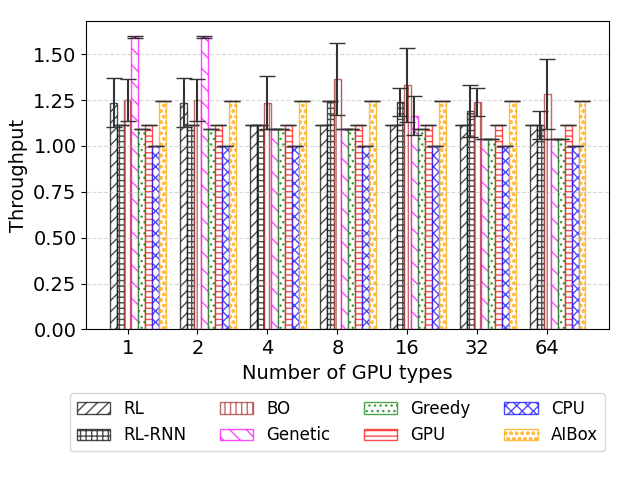}
\vspace{-6mm}
\caption{The normalized throughput corresponding to each scheduling method. The throughput is normalized by dividing the throughput limit for the sake of easy comparison.}
\label{fig:throughputMatchnet} 
\end{figure}

\begin{table}[!t]
\centering
\caption{Scheduling time corresponding to Brute Force (BF) and the Reinforcement learning-based method. The number in the parentheses represent the number of GPU types. We consider the combination of CPUs and diverse types of GPU types. ``E'' represents estimated time because of long scheduling time. As the scheduling time of RL for different numbers of GPU types are similar, we only present the average scheduling time of RL with 2 and 4 GPU types. ``/'' represents that the scheduling time is too long to be compared. The time unit is second.}
\label{tal:schedulingTime}
\begin{tabular}{|c|c|c|c|c|c|}
\hline
Number of layers & BF(2) & BF(4) & RL \\
\hline
8 & 0.068 & 11.37 & 15.91 \\
\hline
12 & 0.71 & 3776.82 & 16.83 \\
\hline
16 & 11.74 & 1254017.23(E) & 17.16 \\
\hline
20 & 217.26 & / & 19.13 \\
\hline
\end{tabular}
\end{table}

As shown in Figure \ref{fig:throughputProvisioning}, we can see that the cost of our provisioning method significantly outperforms the heuristics, i.e., StaRatio (up to 57.9\%) and StaPSRatio (up to 48.3\%).
This result is expected, as our method tries to achieve load balancing among multiple stages and chooses an appropriate number of computing resources for the distributed training process.
In addition, we find that StaPSRatio outperforms StaRatio (up to 55.8\%) as it takes into consideration the number of CPU cores for parameter servers. 

\subsection{Layer Scheduling}
\label{subsec:layerScheduling}

In this section, we present the experimentation to show the advantages of our proposed scheduling methods, i.e., the Reinforcement Learning (RL)-based method with LSTM. We compare RL with six baseline methods, i.e., Brute Force (BF), \rev{RL with an Elman RNN model \cite{vu2016bi} (RL-RNN),} Bayesian Optimization-based method (BO) \cite{Dolatnia2016Bayesian}, CPU, GPU, Genetic \cite{barika2019scheduling}, Greedy \cite{shi2020device}, and Heuristic \rev{(BytePS)} \cite{Zhao2019AIBox}. CPU and GPU represent that the execution of all the layers is carried out in CPUs and GPUs. Heuristic represents that the execution of the first layer is carried out in GPUs and the rest is carried out in CPUs. We take the V100 GPU with different prices to simulate multiple types of GPUs in this section. \revv{In the experimentation, the simulation with the divers GPUs is reflected based on the calculation of the cost. As the GPUs of divers costs may lead to divers costs of the training process, the final scheduling plans can reflect the utilization of divers GPUs.}

\begin{figure}[!t]
\centering
\includegraphics[width=0.48\textwidth]{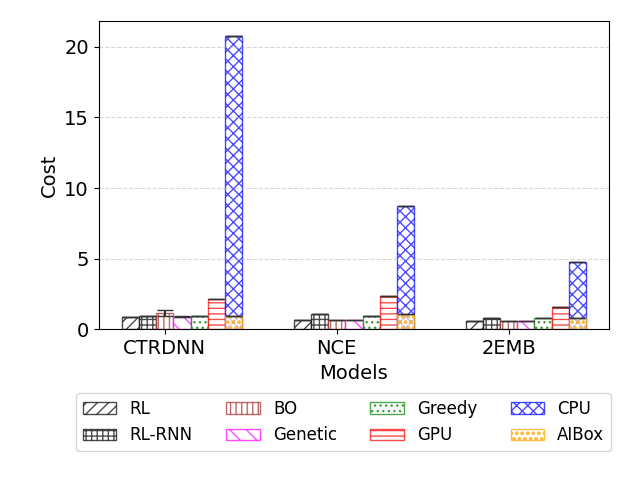}
\vspace{-6mm}
\caption{The cost corresponding to diverse models. The cost is normalized by multiplying by a constant value for the sake of easy comparison.}
\label{fig:costModelsFull} 
\end{figure}

\begin{figure}[!t]
\centering
\includegraphics[width=0.48\textwidth]{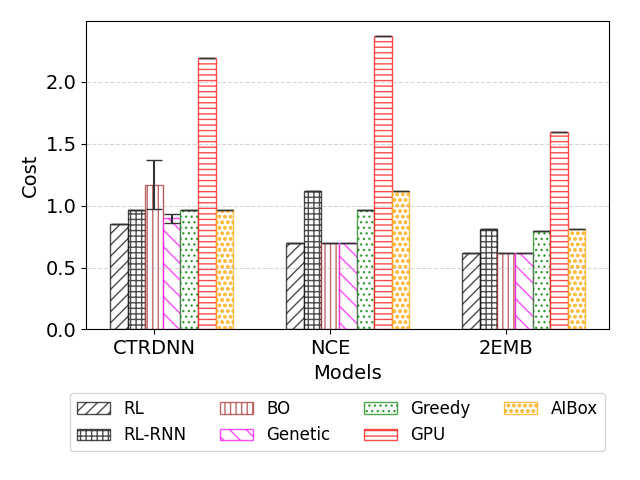}
\vspace{-6mm}
\caption{The cost corresponding to diverse models without CPU. The cost is normalized by multiplying by a constant value for the sake of easy comparison.}
\label{fig:costModels} 
\end{figure}

We focus on four types of models (see the structures in Appendix): MATCHNET (16), CTRDNN (16), 2EMB (10), and NCE (5), with the number of layers in parenthesis. Although MATCHNET and CTRDNN have the same number of layers, MATCHNET is more complex than CTRDNN because of the diverse types of layers. 

First, we modify the structure of CTRDNN by removing or adding Full Connection (FC) layers to simulate a network of 8, 12, 16, and 20 layers. We use a Brute Force (BF) method and the RL method to carry out the scheduling process. We find that the costs of BF and RL are the same, while the scheduling time of BF is much longer than that of RL when the number of layers is big as shown in Table \ref{tal:schedulingTime}. Although it can generate the optimal scheduling plans, BF corresponds to an extremely long scheduling time when the number of layers is big, which is not practical. In addition, we find that the scheduling plans generated by the RL method are the same as the optimal plans generated by BF. 

\begin{figure}[!t]
\centering
\includegraphics[width=0.48\textwidth]{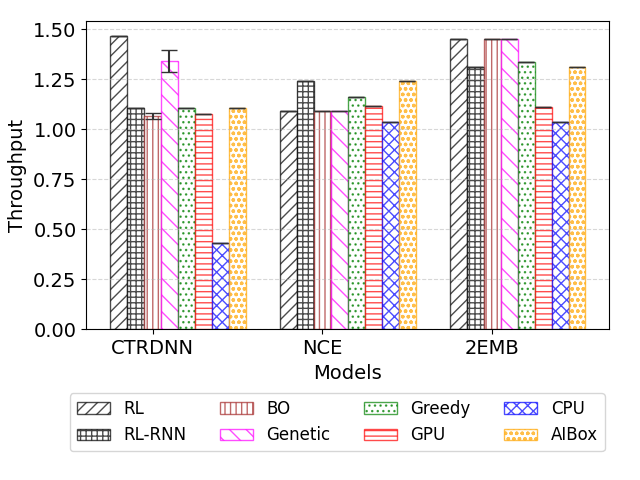}
\vspace{-6mm}
\caption{The throughput corresponding to diverse models. The throughput is normalized by dividing the throughput limit for the sake of easy comparison.}
\label{fig:throughputModels} 
\end{figure}

\begin{figure}[!t]
\centering
\includegraphics[width=0.48\textwidth]{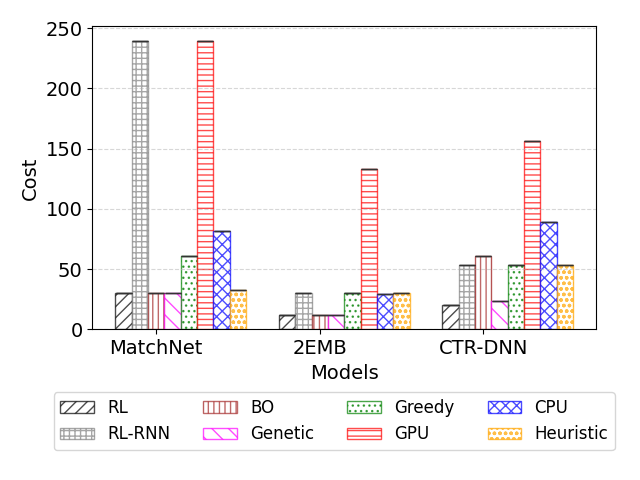}
\vspace{-6mm}
\caption{The cost corresponding to diverse models and scheduling methods based on real execution. The unit of the monetary cost is USD.}
\label{fig:realCost} 
\end{figure}

\rev{
\begin{table*}[!t]
\centering
\caption{Scheduling time corresponding to diverse approaches. The time unit is second. ``32'' represents 32 types of computing resources.}
\label{tal:schedulingTimeModel}
\begin{tabular}{|c|c|c|c|c|c|c|c|c|}
\hline
Model & RL-LSTM & RL-RNN & BO & Genetic & Greedy & GPU & CPU & Heuristic \\
\hline
MATCHNET & 89 & 168 & 1042 & 51 & 0.02 & 0.0003 & 0.0004 & 0.002 \\
\hline
MATCHNET (32) & 62 & 172 & 807 & 68 & 0.27 & 0.0004 & 0.0002 & 0.0004 \\
\hline
MATCHNET (64) & 55 & 164 & 906 & 74 & 0.57 & 0.0004 & 0.0003 & 0.0005 \\
\hline
CTRDNN & 58 & 182 & 602 & 43 & 0.02 & 0.0004 & 0.0002 & 0.0004 \\
\hline
2EMB & 18 & 64 & 693 & 35 & 0.009 & 0.0004 & 0.0002 & 0.0004 \\
\hline
NCE & 12 & 126 & 464 & 23 & 0.004 & 0.0003 & 0.0002 & 0.0004 \\
\hline
\end{tabular}
\end{table*}
}

We calculate the cost based on the simulation with the cost model presented in Section \ref{subsec:costModel} and the throughput and cost measured from the real execution of MATCHNET.
The costs corresponding to diverse scheduling methods while ensuring the throughput are shown in Figures \ref{fig:costMatchnetFull} and \ref{fig:costMatchnet}. In terms of the cost, RL significantly outperforms the baseline methods, i.e., \rev{RL-RNN (up to 321.1\%),} BO (up to 27.9\%), Genetic (up to 289.3\%), Greedy (up to 291.4\%), GPU (up to 304.2\%), CPU (up to 4137.3\%), and Heuristic (up to 312.3\%) \rev{(when the types of resources ranges between 1 and 16). When the scale of the cluster becomes huge and the types of computing resources become significant, the advantages of RL become obvious (%
47.3\% for BO, 1058.0\% for Genetic, 1058.0\% for Greedy, 1757.5\% for GPU, 19370.1\% for CPU, and 1794.7\% for Heuristic with 32 types of resources) and (%
1811.6\% for BO, 1091.2\% for Genetic, 1091.2\% for Greedy, 1819.4\% for GPU, 20019.1\% for CPU, and 2998.5\% for Heuristic with 64 types of resources).}  \rev{As LSTM leverages the forget gate and we design the specific input features for each layer, our approach corresponds to better results compared with RL-RNN, which suffers from the vanishing gradients problem \cite{henaff2016recurrent}.} In addition, as RL with LSTM can learn more information than BO and other heuristics, our approach outperforms other heuristics in most cases. \rev{We find that BO corresponds to higher variance compared with RL with LSTM.} Furthermore, although the price of CPU is much lower than that of GPU, we find that the cost of using CPU for the training is higher than that of GPU in the simulation \rev{as shown in Figure \ref{fig:costMatchnetFull}}. Please note that ``cost'' is calculated based on the price of CPU or GPU and the execution time, which is different from ``price'' (the amount of money for using a computing resource within a time unit). CPU corresponds to bigger numbers of servers to achieve similar throughput as that of other scheduling methods, which incurs high cost.
Figure \ref{fig:throughputMatchnet}\footnote{In a big cluster or grid computing platform, e.g., Grid5000 \cite{grid5000}, we can easily have diverse types (40+) of computing resources.} shows the normalized throughput of all the scheduling methods, which reflects that all the scheduling methods can meet the throughput constraint. 

\begin{figure}[!t]
\centering
\includegraphics[width=0.43\textwidth]{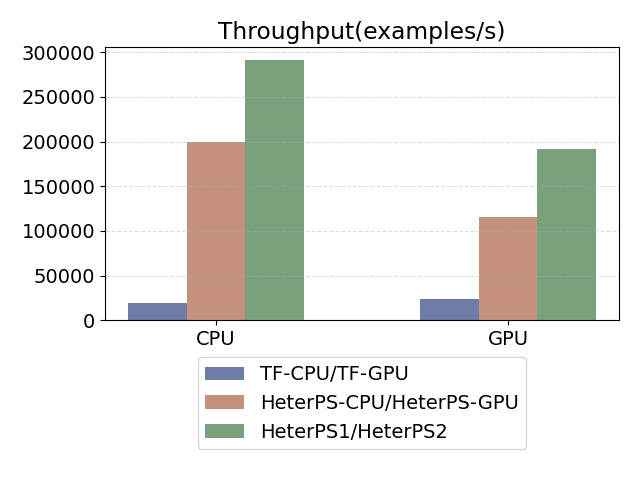}
\caption{The throughput corresponding to TF-CPU, TF-GPU, HeterPS-CPU, HeterPS-GPU, and HeterPS. CPU represents the experiment with CTRDNN1, and GPU represents that with CTRDNN2.}
\label{fig:throughputTF} 
\end{figure}

Figures \ref{fig:costModelsFull} and \ref{fig:costModels} demonstrate the cost to train multiple models with diverse methods. RL significantly outperforms baseline methods, i.e., \rev{RL-RNN (up to 37.3\%),} BO (up to 38.1\%), Genetic (up to 6.2\%), Greedy (up to 29.3\%), GPU (up to 229.0\%) and Heuristic (up to 57.4\%). Although it performs as well as RL for NCE and 2EMB, BO corresponds to a high cost for CTRDNN and has a significant variance, which is incurred by the randomness within the sampling process of BO. In addition, as the structure of CTRDNN is more complex than other networks, it is hard for BO to find a good scheduling plan. Figure \ref{fig:throughputModels} demonstrates the throughput for multiple models with different methods. The throughput of all cases meets the constraint, except for the case of CPU for CTRDNN. This is because the number of CPU servers is restricted by a limit, i.e., there are not enough CPU servers for the training process of CPU to achieve the throughput limit. 



We carry out real execution using CPUs and GPUs. We get similar results, i.e., RL significantly outperforms \rev{RL-RNN (up to 686.2\%),} BO (up to 197.8\%), Genetic (13.46\%), Greedy (159.2\%), CPU (336\%), GPU (1044\%), and Heuristic (159.2\%), as shown in Figure \ref{fig:realCost}. The cost corresponding to CPU significantly differs (up to 17.4 times) from the simulated results. After analysis, we attribute the difference to the overhead brought by small batches within the CPU environments in the profiling information. 
\revv{Although the profiled OCT and ODT may be different from the actual OCT and ODT, the relative values are similar and the results show that RL can significantly outperform other approaches.}
\rev{The throughput limit is selected based on the requirement of the training process allocated for the experimentation. If we take a tighter throughput limit, i.e., higher throughput limit, we may need to provision more resources and the relative advantage of HeterPS may become less significant. Otherwise, the advantage of HeterPS may be come more significant.}

Finally, we measure the scheduling time and the training time of diverse models. 
\rev{The scheduling time of RL for MATCHNET, CTRDNN, 2EMB, NCE is shown in Table \ref{tal:schedulingTimeModel}. Although the RL-LSTM-based scheduling approach takes a longer time to train the model compared with Greedy, GPU, CPU, and Heuristic, it corresponds to smaller cost in diverse environments as shown in Figures \ref{fig:costMatchnet}, \ref{fig:costModels}, and \ref{fig:realCost}. The training time of RL-LSTM-based scheduling approach is much shorter than RL-RNN and BO. When the scale of the computing resource types become significant, the scheduling time of RL-LSTM does not increase.}
In addition, the average training time of MATCHNET, CTRDNN, 2EMB, and NCE is 4980s, 3600s, 2040s, and 5340s, respectively. The time to train the LSTM model ranges from 0.2\% to 1.8\% of the total training time, which is negligible. In addition, the trained LSTM model can be used multiple times with different input data (the inference time is less than 0.002s).

\subsection{Throughput}

To show the efficiency of \TheName{}, we compare the throughput of \TheName{} with that of Tensorflow \cite{abadi2016tensorflow}. We execute a CTRDNN of 7 layers, which corresponds to two models of different dimensions. We take ``CTRDNN1'' to denote the low dimension one and ``CTRDNN2'' to denote high dimension one. We take 4 CPU servers and 4 GPU servers while using the RL scheduling method to carry out the scheduling process. We conduct the training process of CTRDNN1 with Tensorflow of CPU configuration (TF-CPU), \TheName{} with the CPU scheduling method (\TheName{}-CPU), and \TheName{}1~using both CPUs and GPUs (\TheName{}), and that of CTRDNN2 with Tensorflow of GPU configuration (TF-GPU), \TheName{}~with the GPU scheduling method (\TheName{}-GPU), and \TheName{}2~using both CPUs and GPUs (\TheName{}). Figure \ref{fig:throughputTF} shows that HeterPS-CPU (9.5 times), HeterPS-GPU (3.8 times), HeterPS1 (up to 14.5 times), and HeterPS2 (6.9 times) significantly outperforms Tensorflow with the CPU and GPU configuration, respectively, in terms of throughput.

\section{Conclusion}
\label{sec:con}

To efficiently train a DNN model using heterogeneous computing resources, we propose \TheName{}. The functional architecture of \TheName{} consists of three modules, i.e., distributed training, scheduling, and data management. We exploit the parameter server architecture to handle the distributed machine learning. We utilize multiple data management methods to improve efficiency. In addition, we propose a reinforcement learning-based scheduling method to schedule each layer in order to a proper type of computing resource to minimize the cost while ensuring the throughput limit based on a provisioning method. We carried out extensive experimentation, which demonstrates the advantages of our framework. The experimental results show that the provisioning method can be up to 57.9\% better than baseline methods, the scheduling method can significantly outperform state-of-the-art methods (up to 312.3\% in terms of the monetary cost), and the throughput of the framework is 14.5 times more significant than TensorFlow.

\rev{Although HeterPS enables the large-scale training with elastic heterogeneous resources, it cannot handle the decentralized data across multiple data centers and does not consider security and privacy of the decentralized data. In the future, we plan to extend \TheName{} to multiple data centers to process decentralized data. As the privacy and security of some sensitive data are of much importance, some data may be stored and accessed in a specific data center. While \TheName{} can support the distributed training with arbitrary model partition, and is compatible with vertical federated learning \cite{liu2021distributed,liu2023distributed}, we can optimize the system architecture and the scheduling method of \TheName{} to efficiently train a big model while ensuring the security and privacy of the decentralized data. In addition, we may combine the scheduling process and the provisioning process while using a unified RL process to achieve better performance in the future. Furthermore, resource sharing may be exploited in the future to address possible difficulty in balancing the throughput of diverse stages.}

\bibliography{reference}
\bibliographystyle{abbrv}


\clearpage

\appendix
\section*{Appendix}

\begin{figure}[!ht]
\centering
\includegraphics[width=0.48\textwidth]{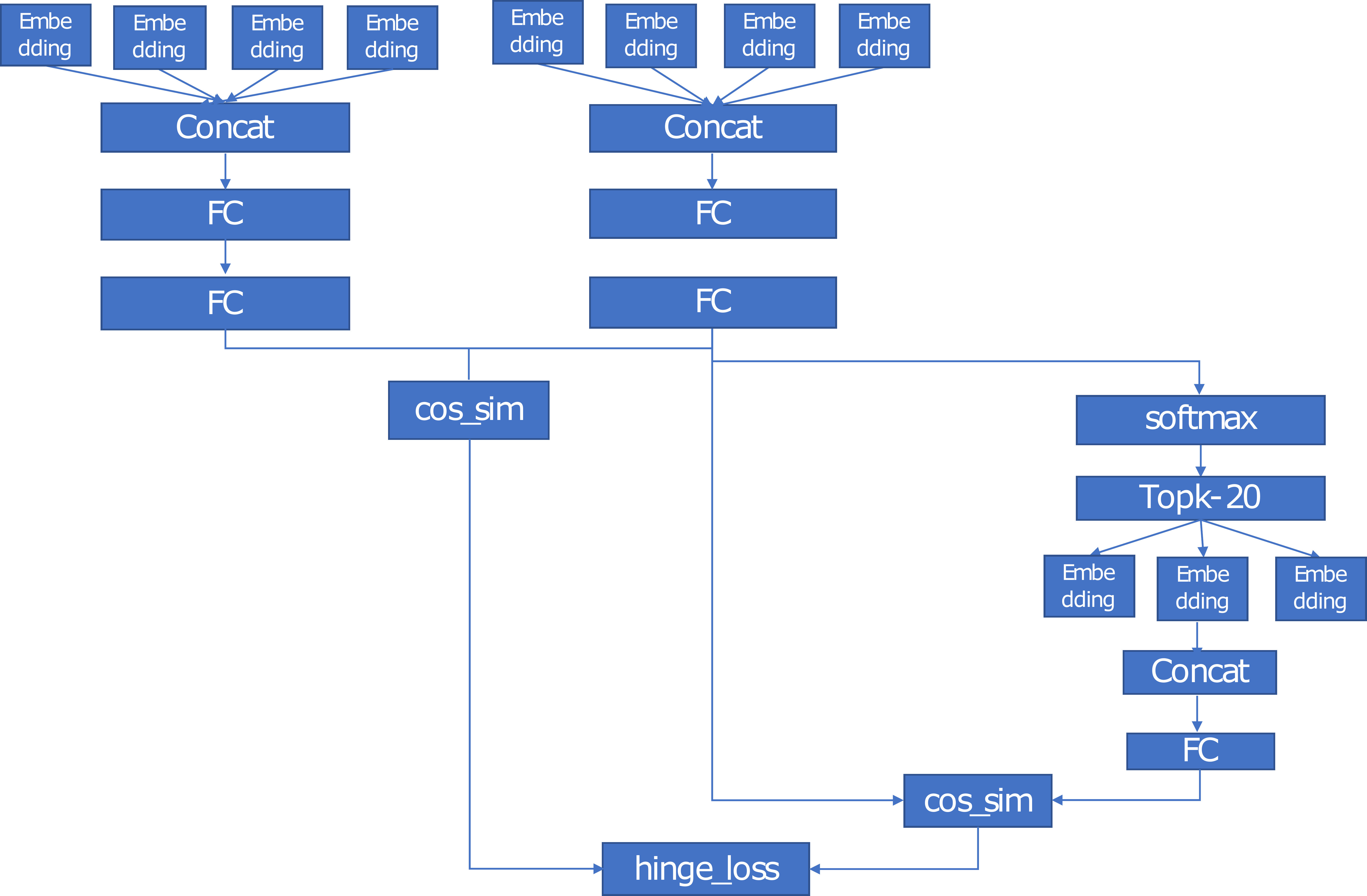}
\caption{The structure of MATCHNET. }
\label{fig:matchnet} 
\end{figure}

\begin{figure}[!ht]
\centering
\includegraphics[width=0.2\textwidth]{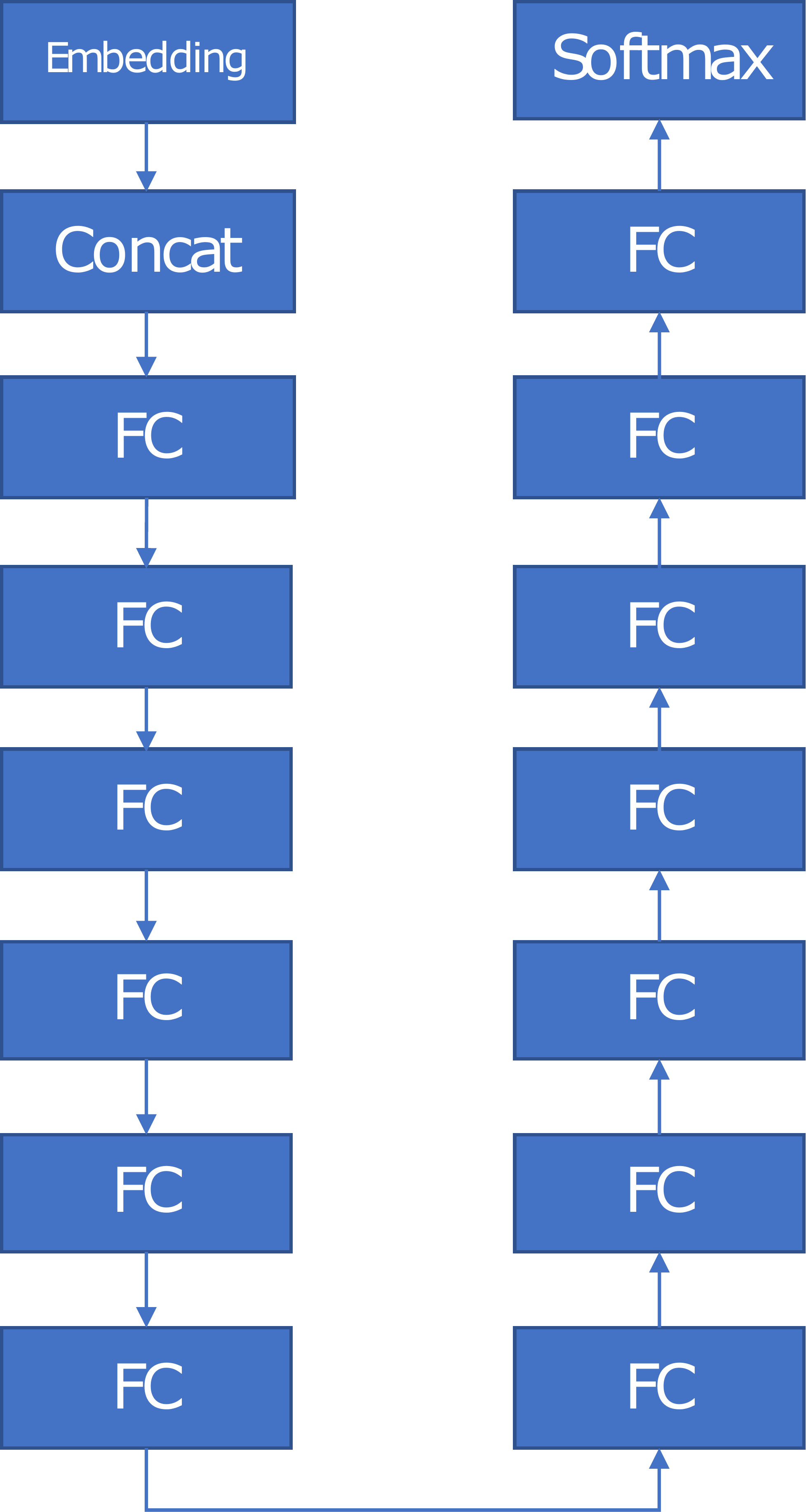}
\caption{The structure of CTRDNN. }
\label{fig:ctrdnn} 
\end{figure}

\begin{figure}[!ht]
\centering
\includegraphics[width=0.4\textwidth]{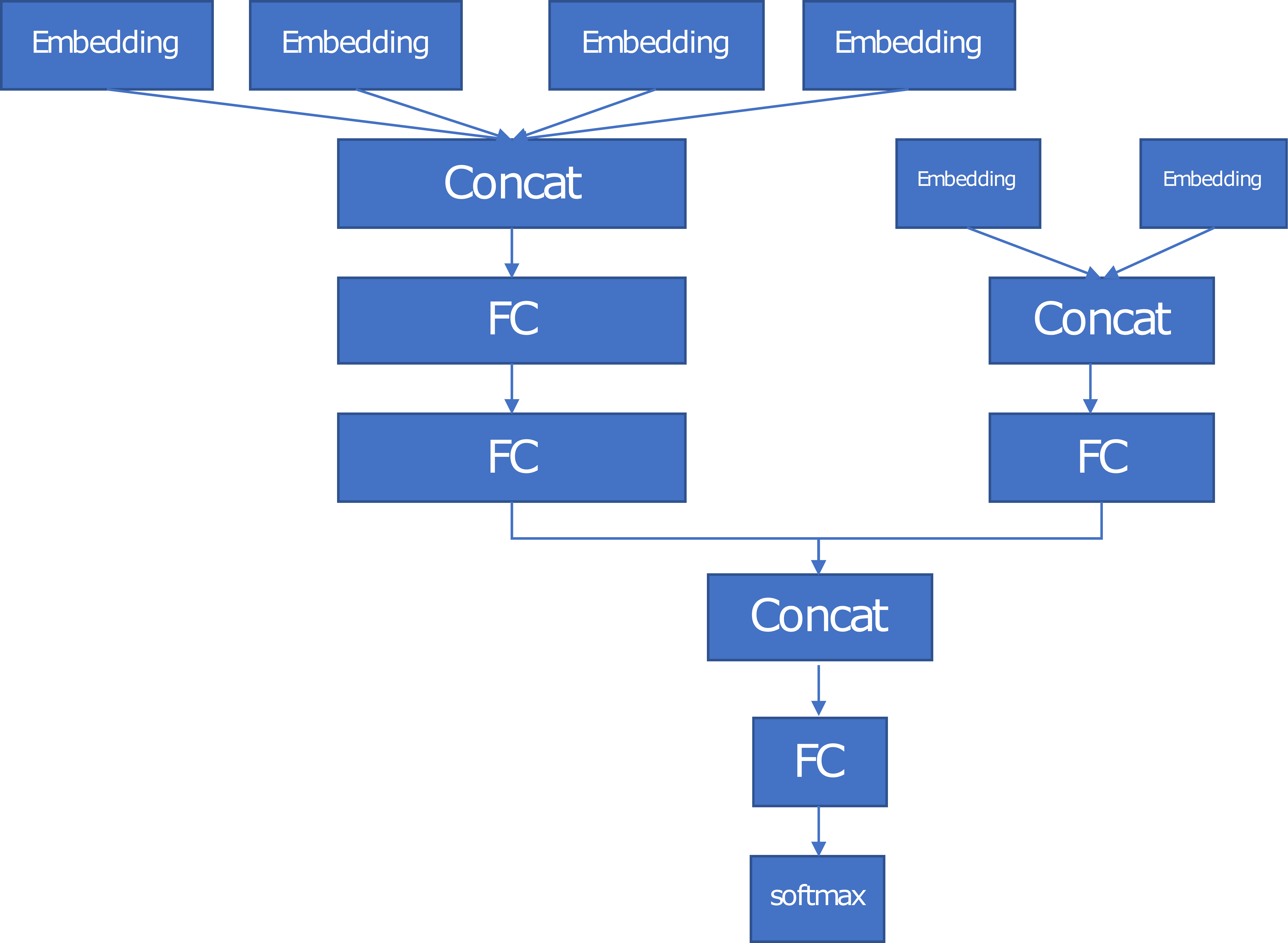}
\caption{The structure of 2EMB. }
\label{fig:2emb} 
\end{figure}

\begin{figure}[!ht]
\centering
\includegraphics[width=0.36\textwidth]{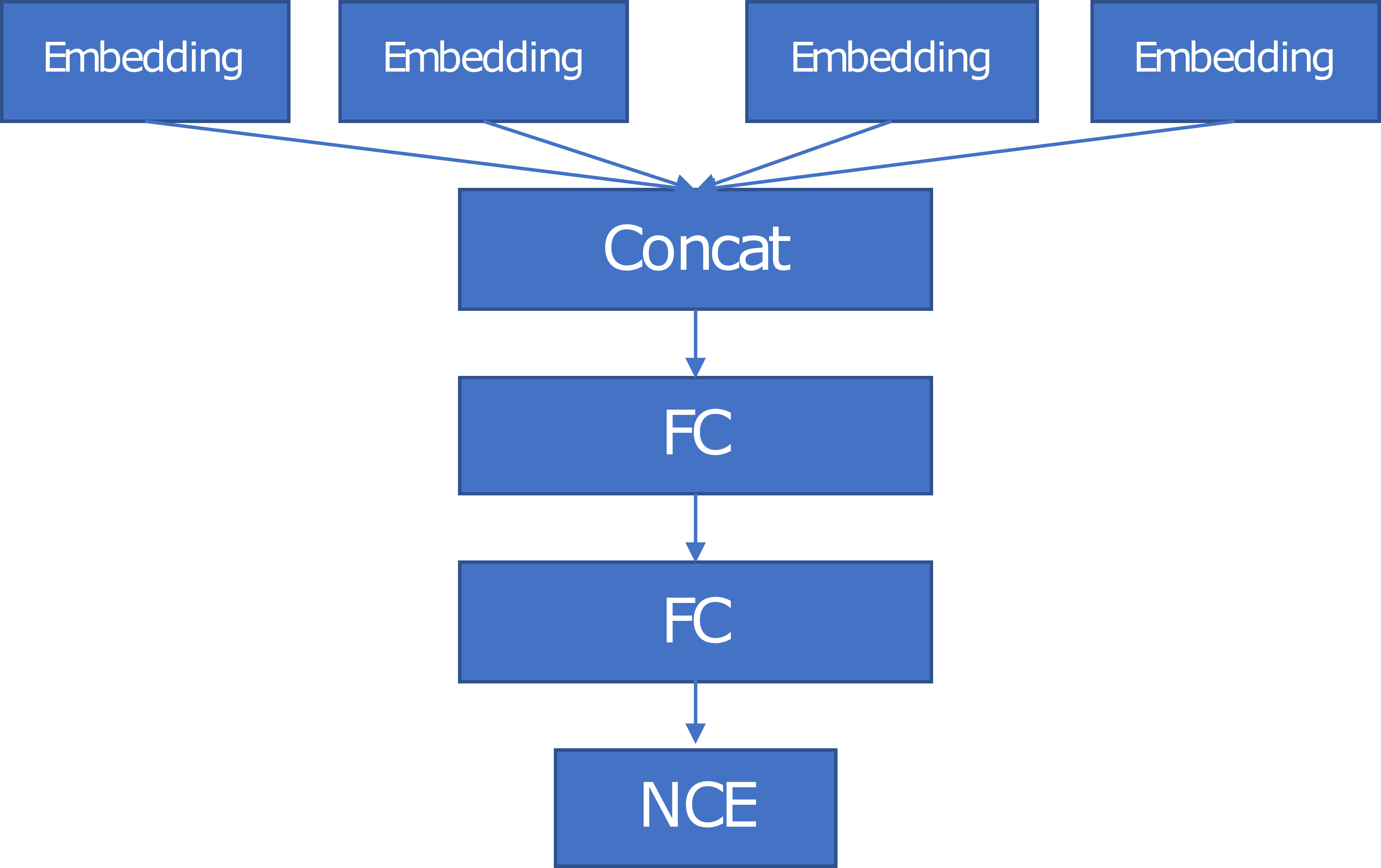}
\caption{The structure of NCE. }
\label{fig:nce} 
\end{figure}

The structures of MATCHNET, CTRDNN, 2EMB, and NCE are shown in Figures 
\ref{fig:matchnet}, \ref{fig:ctrdnn}, \ref{fig:2emb}, \ref{fig:nce}, respectively. 

\end{document}